\documentclass[review,12pt]{elsarticle}

\usepackage[ansinew]{inputenc}
\usepackage{graphicx}
\usepackage{color}
\usepackage[colorlinks]{hyperref}
\usepackage{graphics}
\usepackage{latexsym}
\usepackage{showidx}
\usepackage{amssymb}
\usepackage{amsfonts}
\usepackage{amsmath,amsthm}

\font\msbm=msbm10 at 12pt
\newcommand{\R}{\mbox{\msbm R}}

\newcommand{\Z}{\mbox{\msbm Z}}
\usepackage{amssymb}
\begin{document}

\begin{frontmatter}

%% Title, authors and addresses

%% use the tnoteref command within \title for footnotes;
%% use the tnotetext command for theassociated footnote;
%% use the fnref command within \author or \address for footnotes;
%% use the fntext command for theassociated footnote;
%% use the corref command within \author for corresponding author footnotes;
%% use the cortext command for theassociated footnote;
%% use the ead command for the email address,
%% and the form \ead[url] for the home page:
%% \title{Title\tnoteref{label1}}
%% \tnotetext[label1]{}
%% \author{Name\corref{cor1}\fnref{label2}}
%% \ead{email address}
%% \ead[url]{home page}
%% \fntext[label2]{}
%% \cortext[cor1]{}
%% \address{Address\fnref{label3}}
%% \fntext[label3]{}

\title{Min-Sum algorithm for lattices constructed by Construction $\textit{\textbf{D}}$}

%% use optional labels to link authors explicitly to addresses:
%% \author[label1,label2]{}
%% \address[label1]{}
%% \address[label2]{}

\author{Hassan Mehri}
\ead{hassanmehri.math@gmail.com}
\address{Department of Mathematics and Computer Science, Amirkabir University of Technology, Tehran, Iran}

\begin{abstract}
The so-called min-sum algorithm has been applied for decoding lattices 
constructed by Construction $\textit{\textbf{D}}^{'}$. We generalize 
this iterative decoding algorithm to decode lattices constructed 
by Construction $\textit{\textbf{D}}$. 
An upper bound on the decoding complexity per iteration, in terms of
coding gain, label group sizes of the lattice and other factors
is derived. We show that iterative decoding of LDGM lattices has
a reasonably low
complexity such that lattices with dimensions of a few thousands can be easily decoded.
\end{abstract}

\begin{keyword}
Lattices, Iterative decoding, Min-Sum algorithm, LDGM codes
%%\PACS 89.70.Cf, 89.70.-a, 89.75.Fb, 05.10.-a, 05.45.Tp
%% keywords here, in the form: keyword \sep keyword

%% PACS codes here, in the form: \PACS code \sep code

%% MSC codes here, in the form: \MSC code \sep code
%% or \MSC[2008] code \sep code (2000 is the default)

\end{keyword}

\end{frontmatter}

%% \linenumbers

%% main text
\section{Introduction}
Both of the integer programming method and the trellis approach,
as two main methods for lattice decoding, are impractical in
higher dimensions \cite{Forney1994, Kannan1987}. The min-sum 
algorithm, as an iterative decoding approach, can be used in 
decoding high dimensional lattices.
Tanner generalized Gallager's low-density parity-check (LDPC) codes to 
other class of codes defined by
general bipartite graphs, called Tanner graphs \cite{Tanner1981}. For
any linear binary block code, this construction is based on a
parity check matrix of the code. Tanner graph construction is
used to find a graphical representation of the lattice in terms
of its linear constraints. The decoding complexity of the
generalized min-sum algorithm depends on the Tanner graph
structure and the label code of the lattice. Sadeghi et~al.
introduced a generalization of min-sum decoding algorithm for
lattices constructed by Construction
$\textit{\textbf{D}}^{'}$ \cite{Sadeghi2006}. 
In this work, we will propose
another generalization of min-sum algorithm to decode lattices
constructed by Construction $\textit{\textbf{D}}$. Therefore
properly selected lattices, such as those based on low-density generator 
matrix (LDGM) codes, can
be decoded efficiently. The paper begins in the next section with
a brief discussion about lattice. Section three introduces the
generalized version of min-sum algorithm to decode lattices constructed by
Construction $\textit{\textbf{D}}$. The decoding complexity and it's
bounds for the new generalization of min-sum algorithm are
discussed in the forth section. The final section is dedicated to the paper's 
conclusions.
%-----------------------------------------------
\section{Preliminaries}
Let ${\R}^{m}$ be the $m$-dimensional real vector space with the
standard product $\langle.,.\rangle$ and Euclidean norm
$\parallel\textbf{x}\parallel=\langle\textbf{x},\textbf{x}\rangle^{1/2}$.
A lattice $\Lambda$ is a discrete additive subgroup of ${\R}^{m}$. 
An $n$-dimensional lattice is generated by the integer
combinations of a set of $n$ linearly independent vectors
\cite{Conway1999}. Any subgroup of a lattice $\Lambda$ is
called sublattice of $\Lambda$ and a lattice is called
\textit{orthogonal} if it has a basis with mutually orthogonal
vectors. The set $\Lambda^{*}$ of all vectors in the real span of
$\Lambda$ (span($\Lambda$)), whose the standard inner product with
all elements of $\Lambda$ has an integer value, is an
$n$-dimensional lattice called the \textit{dual} of $\Lambda$.
Let us assume that an $n$-dimensional lattice $\Lambda$ has an
$n$-dimensional orthogonal sublattice $\Lambda^{'}$. If
$\Lambda^{'}$ has a set of basis vectors along the orthogonal
subspaces $S=\{W_{i}\}^{n}_{i=1}$, the projection onto the vector
space $W_{i}$ defined as $P_{W_{i}}$ and the cross section
$\Lambda_{W_{i}}$ defined as $\Lambda_{W_{i}}=\Lambda \cap W_{i}$.
The \emph{label group} of the lattice is defined as
$G_{i}=P_{W_{i}}/\Lambda_{W_{i}}$, which is used to label the
cosets of $\Lambda^{'}$ in $\Lambda$. For any lattice-word
$\textbf{x}$, the label sequence is defined as
$g(\textbf{x})=(g_{1}(\textbf{x}),\ldots,g_{n}(\textbf{x}))$,
where $g_{i}(\textbf{x})=P_{W_{i}}+\Lambda_{W_{i}}$. The set of
all possible label sequences,
$L=g(\Lambda)=\{g(\textbf{x}):\textbf{x}\in\Lambda\}$, is the
label code of $\Lambda$. This set is an Abelian group block code
over the lattice-alphabet sequence space,
$\textbf{G}=G_{1}\times\ldots\times G_{n}$. Let $|G_{i}|=g_{i}$
and $\textbf{v}_{i}$ be the generator vector of $\Lambda_{W_{i}}$,
\textit{i.e.}, $\Lambda_{W_{i}}=\Z \textbf{v}_{i}$. Each element
of $G_{i}$ can be rewritten in the form of
$\Lambda_{W_{i}}+j\det(P_{W_{i}})\textbf{v}_{i}/|\textbf{v}_{i}|(j=1,\ldots,g_{i}-1)$.
Then the map
\begin{equation}
\label{eq01}
\Lambda_{W_{i}}+j\det(P_{W_{i}})\frac{\textbf{v}_{i}}{|\textbf{v}_{i}|}\longrightarrow j
\end{equation}
is an isomorphism between $G_{i}$ and ${\Z}_{g_{i}}$ \cite{Sadeghi2006}, thus every element 
of the label group $G_{i}$ can be written as
$(\Z+a_{j})\textbf{v}_{i}$, where $a_{j}=j~det(P_{W_{i}})/det(\Lambda_{W_{i}})$.

There exist another efficient method for lattice representation
introduced by Tanner \cite{Forney1994}. Lattices constructed by
Construction $\textit{\textbf{D}}$ have a square generator matrix.
If $\textbf{B}$ is a generator matrix for $\Lambda$, then
$\textbf{B}^{*}=(\textbf{B}^{-1})^{tr}$ is a generator matrix for
$\Lambda^{*}$ (parity-check matrix of $\Lambda$)
\cite{Barnes1983}. This can be applied to construct the Tanner
graph for the lattice \cite{Conway1999}. To construct the Tanner
graph for a lattice, The Tanner graph construction of linear
codes is applied to the parity check matrix of the lattice. If
$s_{i}$ denotes the $i$th column and $ch_j$ denotes the $j$th row
of $\textbf{B}^{*}$ respectively, then the Tanner graph of the
lattice has the edge $(s_{i},ch_{j})$ if and only if the
lattice-word component (symbol node) $s_{i}$ is contained in (or
checked by) the parity-check sum (check node) $ch_{j}$. As a
result $\textbf{x}\in{\Z}^{n}$ belongs to $\Lambda$ if
and only if $\textbf{B}^{*}\textbf{x}^{T}\in{\Z}$.\\

\begin{figure}[h]
\centerline{\includegraphics[height=5cm,width=13cm]{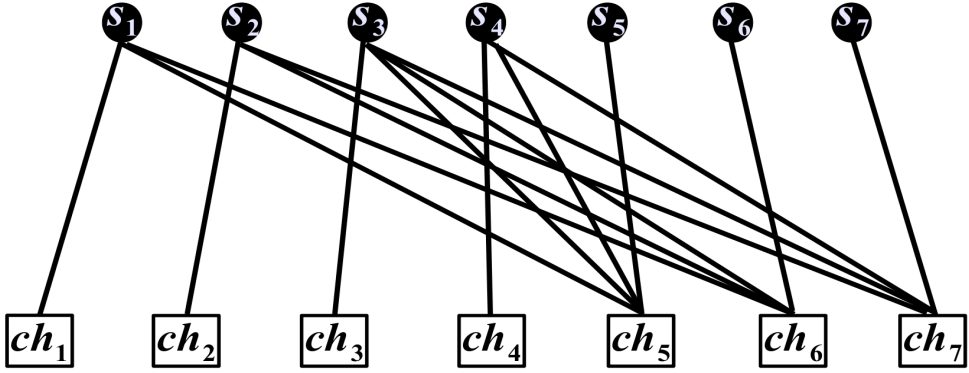}}
\caption{The corresponding Tanner graph for the 7-dimensional lattice discussed in 
example 2.1. The black circles and 
the white rectangles represent symbol nodes and check nodes respectively. 
$s_{i}$ denotes the $i$th column and $ch_j$ denotes the $j$th row of parity-check 
matrix $\textbf{B}^{*}$.}
\label{sch}
\end{figure}

\noindent{\textbf{Example 2.1}} Consider the following
$\textbf{B}$ and $\textbf{B}^{*}$ as the generator matrix and the
parity check matrix of a 7-dimensional lattice constructed by
Construction $\textit{\textbf{D}}$.
\begin{center}
\begin{equation}
\quad\ \mathbf{B}=
\begin{pmatrix}
\: 1 &\: 0 &\: 0 &\: 0 &\: 1 &\: 1 &\: 0 \: \\
\: 0 &\: 1 &\: 0 &\: 0 &\: 0 &\: 1 &\: 1 \: \\
\: 0 &\: 0 &\: 1 &\: 0 &\: 1 &\: 1 &\: 1 \: \\
\: 0 &\: 0 &\: 0 &\: 1 &\: 1 &\: 0 &\: 1 \: \\
\: 0 &\: 0 &\: 0 &\: 0 &\: 2 &\: 0 &\: 0 \: \\
\: 0 &\: 0 &\: 0 &\: 0 &\: 0 &\: 2 &\: 0 \: \\
\: 0 &\: 0 &\: 0 &\: 0 &\: 0 &\: 0 &\: 2 \: \\
\end{pmatrix},
\,\, \mathbf{B}^{*}=
\begin{pmatrix}
\: 1                       &\: 0                        &\: 0                       &\: 0                       &\: 0           &\: 0           &\: 0           \: \\
\: 0                       &\: 1                        &\: 0                       &\: 0                       &\: 0           &\: 0           &\: 0           \: \\
\: 0                       &\: 0                        &\: 1                       &\: 0                       &\: 0           &\: 0           &\: 0           \: \\
\: 0                       &\: 0                        &\: 0                       &\: 1                       &\: 0           &\: 0           &\: 0           \: \\
\frac{\:}{\:}\frac{1}{2} &\: 0                        &\frac{\:}{\:}\frac{1}{2} &\frac{\:}{\:}\frac{1}{2} &\: \frac{1}{2} &\: 0           &\: 0           \: \\
\frac{\:}{\:}\frac{1}{2} & \frac{\:}{\:}\frac{1}{2}  &\frac{\:}{\:}\frac{1}{2} &\: 0                       &\: 0           &\: \frac{1}{2} &\: 0           \: \\
\: 0                      & \frac{\:}{\:}\frac{1}{2} &\frac{\:}{\:}\frac{1}{2} &\frac{\:}{\:}\frac{1}{2} &\: 0           &\: 0           &\: \frac{1}{2} \: \\
\end{pmatrix}
\nonumber
\end{equation}
\end{center}
The corresponding Tanner graph for this example is illustrated in Fig.~\ref{sch}. 
The black circles denote lattice-word components and the white rectangles represent 
parity-check sums.

%-----------------------------------------------

\section{Generalized min-sum algorithm}
In this section, we explain how
generalized min-sum algorithm for lattices constructed by
Construction $\textit{\textbf{D}}'$ can be changed for decoding
lattices constructed by Construction $\textit{\textbf{D}}$.
%-----------------------------------------------
\subsection{Lattice decoding using min-sum algorithm}
Given a vector $\textbf{y}\in\R^{n}$, the lattice decoding
problem is to find a lattice vector $\textbf{x}$, such that
$\parallel\textbf{y}-\textbf{x}\parallel$ is minimized. Let
$\textbf{y}=\sum^{n}_{i=1}\bar{y}_{i}\textbf{v}_{i}$, where
$\bar{y}_{i}={\langle\textbf{y},\textbf{v}_{i}\rangle}/{\langle\textbf{v}_{i},\textbf{v}_{i}\rangle}$.
The goal of the decoding algorithm is to search for the closest
lattice point to the received-word. By definition:
\begin{equation}
\label{eq02} x^{(i)}_{j}:=a_{j}+\lceil\bar{y_{i}}-a_{j}\rfloor,\
i=1,\ldots ,n\quad {\rm and}\quad j=0,\ldots,|G_{i}|-1
\end{equation}
where $\lceil u\rfloor$ is the closest integer to $u$,
$a_{j}=j~{det(P_{W_{i}})}/{det(\Lambda_{W_{i}})}$ and
$x^{(i)}_{j}$ denotes the closest point in $j$th coset of $G_{i}$
at the $i$th coordinate of the lattice-alphabet sequence space.
Then $x^{(i)}_{j}\textbf{v}_{i}$ is the closest vector of $G_{i}$
to $\bar{y}_{i}\textbf{v}_{i}$. The set
$\textbf{x}^{(i)}:=\{x^{(i)}_{j}:j= 0,\ldots,g_{i}-1 \}$ has
$g_{i}$ candidates of the $i$th coordinate of the
lattice-alphabet sequence space for every component of the
received-word. Define the weight as the squared distance between
the elements of $(\Z+a_{j})\textbf{v}_{i}$ and
$\bar{y}_{i}\textbf{v}_{i}$:
\begin{equation}
\label{eq03}
\omega_{y_{i}}(j):=(x^{(i)}_{j}-\bar{y_{i}})^{2}\parallel\textbf{v}_{i}\parallel^{2}.
\end{equation}
Considering the alphabet sequence space
$\Z_{g_1}\times\ldots\times\Z_{g_{n}}$, we rename $j$ to $c_{i}$
where\ $c_{i}\in \Z_{g_{i}}$. The weight of any valid codeword
$\textbf{c}=(c_{1},\ldots,c_{n})\in L$ correspond to lattice-word
$\textbf{x}=(x^{(1)}_{c_{1}},\ldots,x^{(n)}_{c_{n}})\in \Lambda$,
is defined as
\begin{equation}
\label{eq04}
\omega_{\mathbf{y}}(\mathbf{c})=\sum_{i=1}^{n}\omega_{y_{i}}(c_{i}).
\end{equation}
Now the problem is to find the minimum weight, $\min\big(\omega_{\mathbf{y}}(\mathbf{c})\big)$.
%-----------------------------------------------
\subsection{Min-Sum algorithm for Construction $\textbf{D}$ lattices}
This algorithm includes initialization, hard decision, symbol
node operation and check node operation.\\
\\
\emph{Lemma 3.1:} For any received vector
$\textbf{y}=(y_{1},\ldots,y_{n})\in\R^{n}$ we have
\begin{equation}
\label{eq05}
\omega_{y_{i}}(j)=
\Bigg( \det(\Lambda_{W_{i}}) \left\lgroup\Big(\frac{y_{i}}{det(\Lambda_{W_{i}})}
-\frac{j}{g_{i}}\Big)-\Big\lceil\frac{y_{i}}{det(\Lambda_{W_{i}})}-\frac{j}{g_{i}}\Big\rfloor\right\rgroup \Bigg)^2,
\end{equation}
where $i=1, \ldots, n$ and $j=0, \ldots, g_{i}-1$.
\begin{proof}
Proof is given in \cite{Sadeghi2006}.
\end{proof}
\begin{itemize}
\item[1)]\textit{\textbf{Initialization}:} Let
$\textbf{y}=\sum_{i=1}^{n}\bar{y_{i}}\textbf{v}_{i}=\sum_{i=1}^{n}y_{i}\textbf{e}_{i}\in\R^{n}$
be the received vector. An initial weight is assigned to each
node:
\begin{equation}
\label{eq06}
\omega_{y_{i}}=\big(\omega_{y_{i}}(0),\ldots,\omega_{y_{i}}(g_{i}-1)\big)
\end{equation}
{
\item[2)]\textit{\textbf{Iteration}:} In the iteration step all
weights are alternatively updated to find the lattice-word.
\begin{itemize}
\item[2.1)]\textit{\textbf{Symbol node operation}}: The intermediate weight,
   which denotes the symbol-to-check outgoing weight, computed as follows:
\begin{equation}
\label{eq07}
\omega_{y_{i},ch}(k):=\omega_{y_{i}}(k)+\sum_{ch'\in
Q\atop{y_{i}\in ch'\atop ch'\neq ch}}\omega_{ch',y_{i}}(k), \quad 0\leq
k\leq g_{i}-1
\end{equation}
\item[2.2)]\textit{\textbf{Check node operation}:} The intermediate weight,
   which denotes the check-to-symbol outgoing weight, computed as follows:
\begin{equation}
\label{eq08}
\omega_{ch,y_{i}}(k):=\min_{y_{i}\in
Q_{ch}\atop y_{i}=x^{(i)}_{k}}\sum_{y_{i'}\in ch \atop y_{i'}\neq
y_{i}}\omega_{y_{i'},ch}(k'), \quad 0\leq k'\leq g_{i'}-1
\end{equation}
where $Q$ denotes the set of check equations and $Q_{ch}$ denotes
the set of all valid configurations that satisfy the $ch$th check equation.\\
\end{itemize}
}
\item[3)]\textit{\textbf{Termination}:} For every
symbol node, all incoming messages add to its initial weight to
obtain final weight as follows:
\begin{equation}
\label{eq09} F\omega_{y_{i}}(k):=\omega_{y_{i}}(k)+\sum_{ch'\in Q
\atop y_{i}\in ch'}\omega_{ch',y_{i}}(k).
\end{equation}
\end{itemize}
The goal of the min-sum algorithm is to find a vector
$\textbf{x}=(x^{(1)}_{c_{1}},\ldots,x^{(n)}_{c_{n}})$, which
$x^{(i)}_{k}\in \{x^{(i)}_{0},\ldots, x^{(i)}_{g_{i}-1}\}$ and
the index $k$, obtained as follows:
\begin{equation}
k=arg(\min_{{0\leq k\leq g_{i}-1}} F\omega_{y_{i}}(k)).
\end{equation}
The iteration will stop when the selected vector $\textbf{x}$ is a
lattice-word, \textit{i.e.}, $\textbf{x}$ satisfies all parity
check equations or reaches the maximum number of iteration.
%-----------------------------------------------
\section{Decoding complexity}
In each \emph{Termination} step for every lattice-word component
(symbol node) with label group size $g_{i}$, the number of
comparisons is $g_{i}-1$. Therefore the total number of
comparisons is:
\begin{equation}
\label{eq11}
\Big(\sum_{i=1}^{n}g_{i}\Big) - n.
\end{equation}
In each \emph{Iteration} step, the number of operations
for each symbol node and check node computed separately:\\
\\
\emph{a})  For each symbol node and for each edge, the number of
summation is  $g_{i}(d_{y_{i}}-1)$, where $d_{y_{i}}$ denotes the
number of edges for each symbol node. Thus the total number of
summation would be:
\begin{equation}
\label{eq12}
\sum_{i=1}^{n}g_{i}d_{y_{i}}(d_{y_{i}}-1).
\end{equation}
\emph{b})  For each check node, at most,
$g_{i_{1}}\times\ldots\times g_{i_{d_{ch_{i}}}}$ comparisons
should be made. For each outgoing message,
$\omega_{ch_{i},y_{k}}(j)$, and for each $j$, the number of
summation is $d_{ch_{i}}-1$, where $d_{ch_{i}}$ denotes the
number of edges of each check node. Since there are $d_{ch_{i}}$
edges, the number of summation will not exceed
$d_{ch_{i}}(d_{ch_{i}}-1)(g_{i_{1}}\times\ldots\times
g_{i_{d_{ch_{i}}}})$ summations for each check node. Then the
total number of operations in all check nodes in each iteration is
at most:
\begin{equation}
\label{eq13}
\sum_{i=1}^{n}d_{ch_{i}}(d_{ch_{i}}-1)(g_{i_{1}}\times\ldots\times
g_{i_{d_{ch_{i}}}}).
\end{equation}
Eqs.~(\ref{eq11}), (\ref{eq12}) and (\ref{eq13}) show the
dependency of the decoding complexity on the size of label groups.
In each iteration the total number of operations is:
\begin{equation}
\label{eq14}
\Big(\sum_{i=1}^{n}g_{i}+g_{i}d_{y_{i}}(d_{y_{i}-1})+
d_{ch_{i}}(d_{ch_{i}}-1)(g_{i_{1}}\times\ldots\times
g_{i_{d_{ch_{i}}}})\Big)-n.
\end{equation}
The next Corollary follows from counting  the number of operations and inequality
$g_{i}\geq \big(\gamma(\Lambda)\gamma(\Lambda^{*})\big)^{1/2}$,
 where  $\gamma(\Lambda)$, denotes the coding gain of the lattice \cite{Banihashemi2001}.\\
\\
\emph{Corollary 4.1: (Bounds on decoding complexity)} Let
$\Lambda^{*}$ be the dual of $\Lambda$ and
$\gamma=\big(\gamma(\Lambda)\gamma(\Lambda^{*})\big)^{1/2}$. Also
assume that $\Lambda$ has a Tanner graph with $n$ symbol nodes
and $n$ check nodes for which $g_{i}\leq g$, $d_{y}\leq
d_{y_{i}}\leq d^{max}_{y}$ and $d_{ch}\leq d_{ch_{i}}\leq
d^{max}_{ch}$($i=1,\ldots,n$). The upper bound of decoding
complexity per iteration is:
\begin{equation}
\label{eq15}
n\Big(gd^{max}_{y}(d^{max}_{y}-1)+g^{d^{max}_{ch}}d^{max}_{ch}(d^{max}_{ch}-1)+g-1\Big)
\end{equation}
and the lower bound per iteration is:
\begin{equation} \label{eq16} n\Big(\gamma
d_{y}(d_{y}-1)+\gamma^{d_{ch}}d_{ch}(d_{ch}-1)+\gamma-1\Big).
\end{equation}
The proof is a direct consequence of Eq.~(\ref{eq14}) and the fact
that $g_{i}\geq \gamma$ which has been shown in
\cite{Banihashemi2001}. \\
\\
This corollary shows that the decoding complexity per iteration,
grows linearly with the lattice dimension, $n$, but has power law
dependence on the check nodes degree.

%----------------------------------------

\section{Conclusion}
In this work the min-sum algorithm is generalized to decode
lattices constructed by Construction $\textit{\textbf{D}}$. It is
shown that the upper and lower bounds of decoding complexity
depends on lattice parameters like label group sizes, coding gain,
check nodes and symbol nodes degree of Tanner graph. It is also
shown that the decoding complexity grows linearly with the lattice
dimension, $n$, but has the power law dependence on the check nodes
degree. The analysis of decoding complexity confirms the
usefulness of LDGM codes for the lattice construction. 
It is worth mentioning that the presented decoding algorithm can be used to
decode other constructions of lattices from linear codes.
%---------------------------------
\bibliographystyle{}

\end{document}